\documentclass[10pt,a4paper]{article} % -*- latex -*-
\usepackage[margin=1.5cm]{geometry}
\usepackage[numbers]{natbib}
\usepackage{amsmath}
\usepackage{amsfonts}
\usepackage{paralist}
\usepackage{tikz}
\usepackage[hidelinks]{hyperref}
\hypersetup{colorlinks,linkcolor={blue},citecolor={blue},urlcolor={blue}}
\usepackage[utf8]{inputenc}
\usetikzlibrary{matrix}

\renewcommand{\Re}[0]{\mathbb{R}}
\DeclareMathOperator{\dc}{\mathcal{D}} % derivative change
\DeclareMathOperator{\dcb}{\mathcal{\hat{D}}} % derivative change bundled
\DeclareMathOperator{\diff}{\mathbf{diff}} % one-dimensional derivative
\DeclareMathOperator{\coeff}{\mathbf{coeff}}
\DeclareMathOperator{\FV}{FV}

\title{Evolving the Incremental $\lambda$ Calculus into a Model of Forward AD%
\thanks{\textbf{Extended abstract presented at the AD 2016 Conference, Sep 2016, Oxford UK.}}}

\author{\href{https://www.linkedin.com/in/robert-kelly-61952155}{Robert
    Kelly}\footnote{Corresponding Author, Dept of Computer Science, National
    University of Ireland Maynooth, funded by the Irish Research Council,
    \href{mailto:rob.kelly@cs.nuim.ie}{\texttt{rob.kelly@cs.nuim.ie}}}
  \qquad \href{http://barak.pearlmutter.net}{Barak A.
    Pearlmutter}\footnote{Dept of Computer Science, National University of
    Ireland Maynooth,
    \href{mailto:barak@pearlmutter.net}{\texttt{barak@pearlmutter.net}}}
  \qquad
  \href{http://engineering.purdue.edu/~qobi}{Jeffrey Mark
    Siskind}\footnote{School of Electrical and Computer Engineering, Purdue
    University, \href{mailto:qobi@purdue.edu}{\texttt{qobi@purdue.edu}}}}
\date{April 2016}

\begin{document}
\maketitle
\thispagestyle{empty}

%% Power series extraction machinary.
% coeff i \e. PS_{e} -> coeff i of PS
%% "Obeys" theorems
% Make sure delta R obeys
% PS obey change set axioms
% Propogation of PS through, higher order terms don't affect lower order terms
% Truncation is possible
%% Change "plus" and "times"
% Plus: R -> ZPS_{e} -> PS_{e}
% Times: E -> PS_{e} -> ZPS_{e}
%% Communicative diagram, ILC top left.
% Down 1 -> Power series in epsilon
% Down 2 -> Pwoer series in epsilon truncated at e^2
% Move right -> Uncurry, applies to all levels

\section*{Introduction}

Formal transformations somehow resembling the usual derivative are surprisingly common in computer science, with two notable examples being derivatives of regular expressions \citep{Janusz1964} and derivatives of types \citep{McBride-2001a, Abbott-etal-2004a}.
A newcomer to this list is the incremental $\lambda$-calculus, or ILC, a ``theory of changes'' that deploys a formal apparatus allowing the automatic generation of efficient update functions which perform incremental computation \citep{Cai:2014:TCH:2594291.2594304}.
An example of this would be using the ILC derivative-like operator $\dc$ to alter a function $f: B \rightarrow B$, which performs some major reorganization on a database (of type $B$), into the update function $\dc f: B \rightarrow \Delta B \rightarrow \Delta B$.
Here $\Delta B$ is the type of \emph{changes} to $B$.
So $\dc f$, given an initial database, maps a \emph{change} to that input database to a change to the output database.
This in principle, and as shown in their work also in practice, allows enormous savings when the change to the input is small compared to the size of the input itself.
Resemblance to the standard derivative can be exhibited by a simple example
\begin{subequations}
  \begin{align}
    \dc \: (\lambda \; x \; . \; f \; (g \; x))
    &\leadsto (\lambda \; x \; x' \; . \; \dc f \; (g \; x) \; (\dc g \; x \; x'))
    \\
    \intertext{or}
    \dc \: (f \circ g) \; x
    &\leadsto \dc f \; (g \; x) \circ \dc g \; x
  \end{align}
\end{subequations}
which seems suspiciously similar to the familiar Calculus~101 chain rule.

The ILC is not only defined, but given a formal machine-understandable definition---accompanied by mechanically verifiable proofs of various properties, including in particular correctness of various sorts.
Here, we show how the ILC can be mutated into propagating tangents, thus serving as a model of Forward Accumulation Mode Automatic Differentiation.\footnote{%
The approach detailed here stands in contrast to the Simply Typed $\lambda$-Calculus of Forward Automatic Differentiation \citep{Manzyuk-mfps2012}.
Aside from some issues with confluence, that work folded together levels of the hierarchy by not distinguishing numeric basis functions which operate on $\Re$ from those which are lifted to operate on Dual numbers, while here these are distinguished. Moreover, here we have a framework for machine-readable machine-verified proofs of various correctness and efficiency properties.
This approach differs from the Differential Lambda-Calculus \citep{Ehrhard-Regnier-2003a} in analogous ways: complexity, machine-checked proofs, and explicit segregation of levels of differentiation.}

This mutation is done in several steps.
These steps can also be applied to the proofs, resulting in machine-checked proofs of the correctness of this model of forward AD.

\section*{The Mutagenic Steps}

There are two differences between the incremental $\lambda$ calculus and forward AD.
First, \emph{changes} rather than \emph{tangents} are propagated.
These changes are elements of \emph{change sets}, and constitute finite (i.e., not infinitesimal) modifications.
(For example, a change to a list might consist of swapping the first two elements, and a change to a number might consist of increasing its value by 5.)
In numerics, these would be ``differences'' rather than ``differentials'', and $\Delta$ rather than $\partial$.
Second, the changes are passed as additional arguments instead of being bundled together with primal values.
Passing changes in additional arguments makes great sense in the domain of incremental computation, where the whole point of the construction is to partially evaluate a function $\dc f : \alpha \rightarrow \Delta \alpha \rightarrow \Delta \beta$ with respect to $f$'s original input, yielding a mapping of changes to changes: $ \Delta \alpha \rightarrow \Delta \beta$.
But in the context of forward AD, we wish to propagate tangent values \emph{in parallel} with primal values, which necessitates both bundling the ``new'' values with the original ones, and including the original output in the output of the transformed function.

\begin{figure}[t!]
\begin{center}
% Lovely commutative diagram! TODO : Make transitions and states intelligent.
\begin{tikzpicture}
  \matrix (m) [matrix of math nodes,row sep=20ex,column sep=10em,minimum width=2em]
  {
    \text{\shortstack{Incremental\\$\lambda$-Calculus}}	& \text{\shortstack{Power\\Series}}		& \text{\shortstack{Dual\\Numbers}}	\\
    \text{\shortstack{Bundled\\ILC}}			& \text{\shortstack{Higher-Order\\Forward AD}}	& \text{\shortstack{Forward\\AD}}	\\
};
 \path[-stealth]
   (m-1-1) edge node [above] {$\Delta\Re=\Re[\varepsilon]$} (m-1-2)
           edge node [left] {\rotatebox{-90}{uncurry}} (m-2-1)
   (m-1-2) edge node [left] {\rotatebox{-90}{uncurry}} (m-2-2)
   			edge node [above] {$\Delta\Re=\Re[\varepsilon]/\varepsilon^2$} (m-1-3)
   (m-2-1) edge node [above] {$\Delta\Re=\Re[\varepsilon]$} (m-2-2)
   (m-2-2) edge node [above] {$\Delta\Re=\Re[\varepsilon]/\varepsilon^2$} (m-2-3)
   (m-1-3) edge node [left] {\rotatebox{-90}{uncurry}} (m-2-3);
\end{tikzpicture}
\end{center}
\vspace{-3ex}
\caption{Mutating the Incremental $\lambda$-Calculus (ILC) into Forward-Mode Automatic Differentiation (Forward AD).} \label{fig:diagram}
\end{figure}
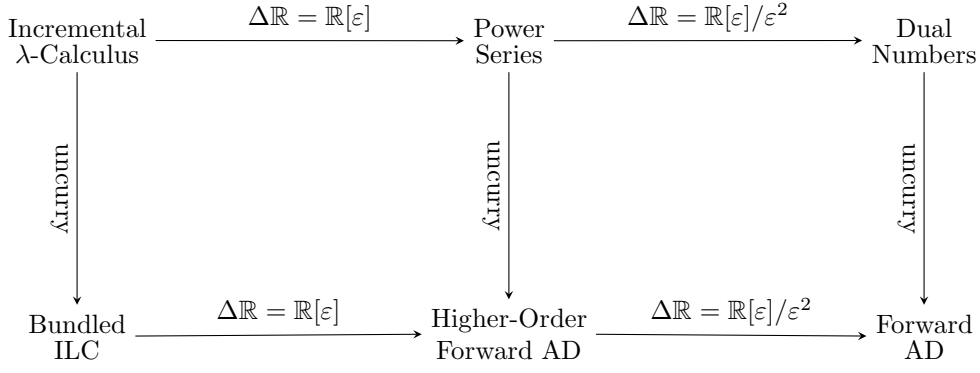

We proceed to eliminate these two differences. This is done in two stages. First, considering only \emph{power series} change sets to the base type $\Re$. And second, \emph{uncurrying} the outputs of the derivative operator and causing it to propagate change sets and primal values bundled together all the way through to its output.
Truncating the power series changes them into Dual Numbers \citep{Clifford1873}, yielding the familiar Forward AD.
A commutative diagram of these steps is shown in Figure~\ref{fig:diagram}. The original ILC is in the top left, with relevant changes indicated with transitions to new states or nodes. Each of these edges leads to a different combination of forward AD in the ILC.  The \emph{power series} and \emph{uncurry} steps can be taken in either order, so the diagram should commute.

Let us describe these two steps in a bit more detail.

\subsection*{Step One: Power Series}

To see how power series change sets are introduced, we note that the ILC allows change sets to be defined for any base type $\tau$.
These change sets need only obey a particular set of axioms, which in our context amounts to associativity of addition of real numbers.
We constrain ourselves to consider only change sets to reals: the base type $\Re$.
We then represent these change sets not as differences, but instead as power series (in some variable $\varepsilon$) with a zero constant term.
\newcommand{\zps}[1]{\langle\textbf{zps}_{#1}\rangle}
\newcommand{\ps}[1]{\langle\textbf{ps}_{#1}\rangle}
This means that the change set of $x:\Re$ is a term of the form $\zps{\varepsilon}$, where
\begin{subequations}
\begin{align}
  \zps{\varepsilon} &::= 0 \; | \; \varepsilon * \ps{\varepsilon} \\
  \ps{\varepsilon} &::= \Re \; | \; \Re + \zps{\varepsilon} \\[1ex]
  \Delta \Re &\equiv \zps{\varepsilon}
\end{align}
\end{subequations}
For a specific value of $\varepsilon$ (possibly subject to conditions of convergence) this would take on a particular numeric value.
\begin{subequations}
We further define an operator $\coeff$ which takes a nonnegative integer index and a power series in $\varepsilon$ wrapped in a $\lambda$ expression, i.e., $(\lambda \varepsilon \; . \; \ps{\varepsilon})$, and yields the requested coefficient of the given power series.
\begin{align}
 \coeff \; 0 \; (\lambda \varepsilon \; . \; r) &\leadsto r & \text{(where $\varepsilon \not\in \FV(r)$)} \\
 \coeff \; 0 \; (\lambda \varepsilon \; . \; r + \varepsilon * e) &\leadsto r & \text{(where $\varepsilon \not\in \FV(r)$)} \\
 \coeff \; 0 \; (\lambda \varepsilon \; . \; \varepsilon * e) &\leadsto 0 \\
 \coeff \; i \; (\lambda \varepsilon \; . \; r + \varepsilon * e)
 &\leadsto
 \coeff \; (i-1) \; (\lambda \varepsilon \; . \; e) & \text{(where $i>0$ and $\varepsilon \not\in \FV(r)$)} \\
 \coeff \; i \; (\lambda \varepsilon \; . \; \varepsilon * e)
 &\leadsto
 \coeff \; (i-1) \; (\lambda \varepsilon \; . \; e) & \text{(where $i>0$)}
\end{align}
For instance,
\[
\coeff \; 2 \; (\lambda \varepsilon \; . \; 0.1 + \varepsilon * (0.2 + \varepsilon * (0.3 + \varepsilon * (0.4 + \varepsilon * (0.5 + \cdots))))) \leadsto 0.3
\]
\end{subequations}

Useful properties of such a change set are straightforward to establish: closure under the derivatives of the numeric basis functions, and dependence during such operators of coefficients only on coefficients of the same or lower order.
The first property is necessary for consistency, while the second allows these power series to be truncated at $\varepsilon^2$, thus yielding the tangents of standard forward AD.
With this machinery, we could define the familiar derivative $\diff : (\Re\rightarrow\Re)\rightarrow(\Re\rightarrow\Re)$, for instance $\diff \; \sin = \cos$, as
\begin{equation}
  \diff \; f \; x \equiv \coeff \; 1 \; (\lambda \varepsilon \; . \; (\dc \; f \; x \; (\varepsilon * 1)))
\end{equation}
\begin{subequations}
By defining $\coeff$ to distribute over algebraic datatypes
\begin{equation}
  \coeff \; i \; (\lambda \varepsilon \; . \; \textbf{Constructor} \; e_1 \; \cdots \; e_n) \leadsto \textbf{Constructor} \; (\coeff \; i \; (\lambda \varepsilon \; . \; e_1)) \; \cdots \; (\coeff \; i \; (\lambda \varepsilon \; . \; e_n))
\end{equation}
and post-compose over functions
\begin{align}
  \coeff \; i \;  (\lambda \varepsilon \; . \; (\lambda x \; . \; e))
  &\leadsto (\lambda x \; . \; \coeff \; i \; (\lambda \varepsilon \; . \; e))
  & \text{(where $x \not= \varepsilon$)}
\end{align}
this machinery can find directional derivatives of functions with non-scalar output, including Church-encoded output.
\end{subequations}

In this formulation, the tagging necessary to distinguish distinct nested invocations of derivative-taking operators \citep{SiskindPearlmutter2008a, manzyuk-etal-amazing-2015} is handled by the standard $\lambda$-calculus mechanisms for avoiding variable capture during $\beta$-substitution, e.g., $\alpha$-renaming.

\subsection*{Step Two: Uncurrying and Bundling}

The second step is uncurrying arguments, and bundling the output.
We need to change the type of the derivative operator from
\begin{align}
\dc : (t_1 \rightarrow t_2 \rightarrow \cdots \rightarrow t_n \rightarrow u) &\rightarrow (t_1 \rightarrow \Delta t_1 \rightarrow t_2 \rightarrow \Delta t_2 \rightarrow \cdots \rightarrow t_n \rightarrow \Delta t_n \rightarrow \Delta u)
\\
\intertext{to}
\dcb : (t_1 \rightarrow t_2 \rightarrow \cdots \rightarrow t_n \rightarrow u) &\rightarrow (F t_1 \rightarrow F t_2 \rightarrow \cdots \rightarrow F t_n \rightarrow F u)
\end{align}
where $F t$ is isomorphic to $t \times \Delta t$, a primal value bundled with its change set.
If we define $F (t_1 \rightarrow t_2) = F t_1 \rightarrow F t_2$ then this yields a simpler type signature,
\begin{equation}
   \dcb : t \rightarrow F t
\end{equation}
The mechanics of this change are straightforward, requiring that the ILC reductions be modified to take the new shape.
Note that, thus uncurried and carrying primal and change set values in tandem, the chain rules of Equation~1 are simplified: $\dcb \; (f \circ g) \leadsto \dcb \; f \circ \dcb \; g$.

\section*{Acknowledgments}
This work was supported, in part, by Science Foundation Ireland grant
09/IN.1/I2637 and by NSF grant 1522954-IIS.\@
Any opinions, findings, and conclusions or recommendations expressed in this
material are those of the authors and do not necessarily reflect the views
of the sponsors.

\bibliographystyle{unsrtnat}
\begin{small}
  \setlength{\bibsep}{0.2ex}
  \bibliography{ad2016e,../sty/ad,../sty/QobiTeX}
\end{small}

\end{document}